# IRE – 1α may be causing abnormal loss of p53 at post-transcriptional level in chronic myeloid leukemia

Katte Rao Toppaldoddi[1,2]
[1]Hôpital Saint-Louis, Institut Universitaire d'Hématologie, Université Paris Diderot, Paris, France and [2]St. John's Medical Hospital and College, Bangalore, India.
Correspondence (E-mail): katte.rao@gmail.com

**Current treatment strategy for chronic myeloid leukemia (CML) mainly includes inhibition of tyrosine kinase activity, which has dramatically improved the prognosis of the disease but without cure[1]. In addition some patients may become drug-resistant[1]. Thus there is still the need for other therapies to avoid resistance and if possible to cure the disease[1]. Loss of p53 is known to play an important role in the disease progression of CML and causes drug resistance[2]. Here, I propose that in CML, inositol requiring enzyme – 1 alpha (IRE – 1α) may cause abnormal degradation of p53 mRNA resulting in inhibition of apoptosis in leukemic clonal cells, which has not been elucidated before. Hence, I propose that inhibition of endoribonuclease activity of IRE – 1α with small molecule inhibitors may provide a novel strategy to enhance p53 function in CML leukemic clones to overcome the limitations of current treatment regimens.**

CML is a malignant disorder arising from a translocation occurring in a hematopoietic stem cell, which leads to the Bcr – Abl fusion gene[3] encoding the protein p210$^{BCR/ABL}$, a constitutively active form of Abl tyrosine kinase[1]. CML phenotype is very strongly related to the dramatic expansion of hematopoietic progenitors[4] and the progression to acute leukemia in a few years. Concerning to the current therapeutic outcomes, majority of the patients require life – long therapeutic intervention, which permits a clinical, hematological and molecular remission avoiding the progression to acute leukemia. However long term treatment with tyrosine kinase inhibitors may induce non – hematological toxicities. In addition, in around 5% of patients, progression may occur under tyrosine kinase inhibitor treatment. In these cases, allogeneic bone marrow transplantation is the only curative alternative, but requires an HLA matching donors and is associated with an important morbidity.

In CML, loss of p53 function is known to induce drug resistance and aggressive progression to blast crisis[2]. But it is not known that even though when the p53 allele is unmutated and exists with intact gene regulatory elements how loss of p53 mRNA expression occurs during disease progression of CML[2, 5, 6, 7]. It has been suggested that downregulated p53 protein expression in CML could be due to either destabilization of its mRNA or epigenetic modulation[5]. Further, it has been demonstrated that down – regulated p53 protein expression in CML is not due to defective epigenetic modulation or defective transcriptional factor function responsible for activation of p53 gene expression[7], which suggests that down – regulation of p53 function in CML is possible through destabilization of p53 mRNA also.

Previously it has been demonstrated that, unfolded protein response (UPR) pathway is a downstream target of Bcr – Abl and that UPR confers anti – apoptotic response[8] which indicates that UPR may be aberrantly activated in CML leukemic clonal cells. IRE – 1α is an evolutionarily conserved member of UPR pathway and is found to be over – expressed in Bcr – Abl expressing cells[8]. From the known findings, IRE – 1α has mainly two functions, endoribonuclease and serine/threonine kinase activities. IRE – 1α can be activated by either ER stress or over – expression[9]. IRE – 1α is capable of degrading the target mRNAs by a mechanism called Regulated Ire – 1 Dependent Decay (RIDD)[9]. IRE – 1α induces mRNA degradation by recognising the CUGCAG consensus element accompanied by stem – loop structure in target mRNAs[10].

In the analysis, I found that in the p53 mRNA, IRE – 1α recognition consensus element CUGCAG is accompanied by stem – loop structure. This stem – loop structure is formed from the residue 2218 to 2231 and the recognition consensus element exists from residue 2224 to 2229 in the p53 mRNA. The difference in Gibbs free energy before and after formation of this stem – loop structure is -3.5 units of Gibbs free energy for 14 nucleotide chain calculated by Mfold web server[14] indicating that the formation of this stem – loop structure in the p53 mRNA is spontaneous thus its formation is highly favourable thermodynamically. Hence, the demonstration that inhibition of IRE – 1α diminishes anti – apoptotic response in CML leukemic clonal cells[8] could be mainly due to increased p53 protein expression. Therefore, I propose that inhibition of endoribonuclease activity of IRE – 1α with small molecule inhibitors to enhance p53 function in CML leukemic clones as a novel strategy to overcome limitations of current treatment regimens.

**Results and model for the mechanism:**

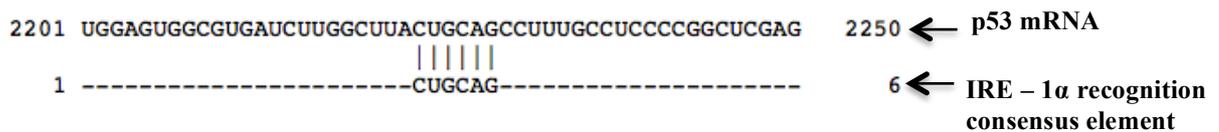

Figure 1 : Nucleotide alignment of p53 mRNA and IRE – 1α recognition consensus element (CUGCAG) using EMBOSS Needle[11, 12, 13]

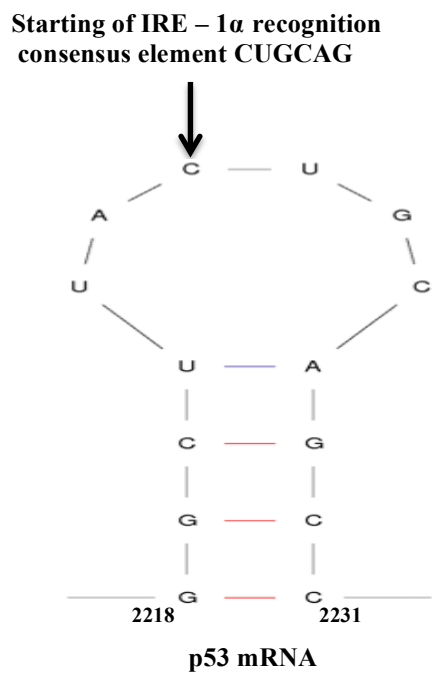

Figure 2 : IRE – 1α recognition consensus element CUGCAG accompanied by stem – loop structure in the p53 mRNA determined by using Mfold Web Server[14]

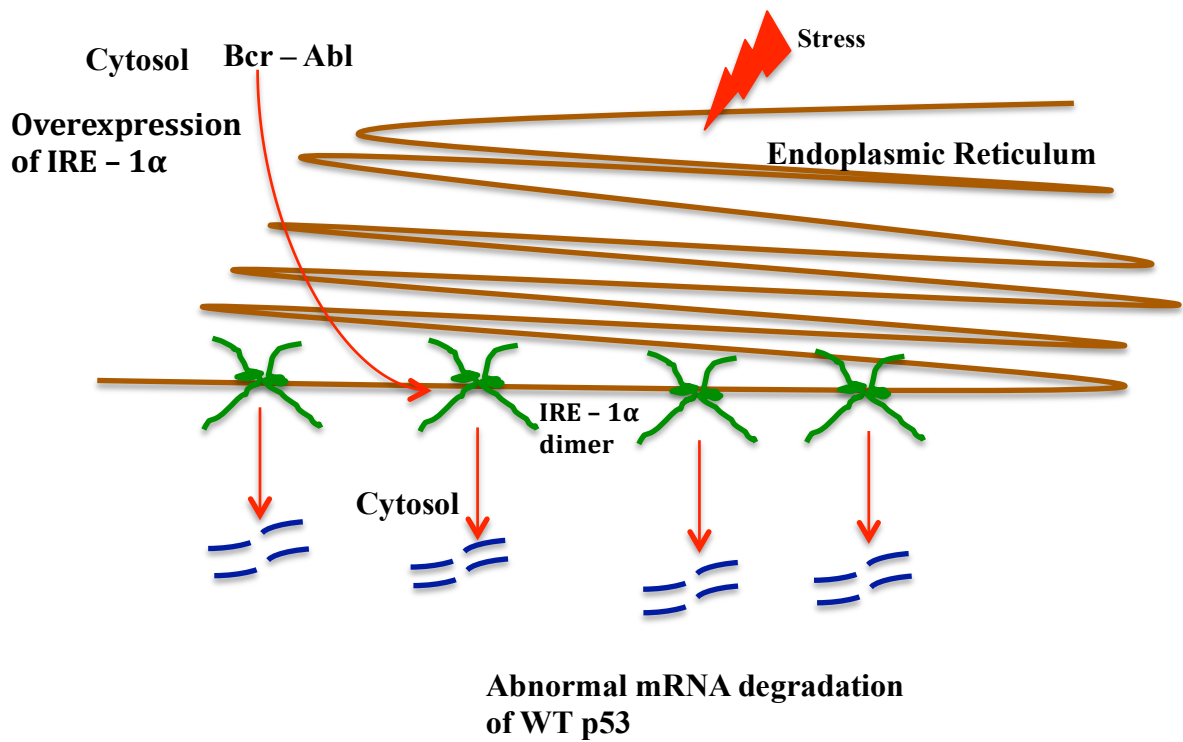

**Figure 3 : Schematic diagram of the model for loss of p53 protein expression in CML**

**Acknowledgement**:

I am thankful to Dr. Sweta Srivastava (St. John's Medical Hospital, India) for the critical discussion. I was supported by fellowships from L'Institut national de la santé et de la recherche médicale (INSERM), France, GLUE Grant Initiative, Department of Biotechnology, Governement of India and travel scholarship from the Service for Science and Technology at the French Embassy (SST), New Delhi, India.